# How much gallium do we need for a p-type Cu(In,Ga)Se$_2$?

Omar Ramírez[1*], Evandro Martin Lanzoni[1], Ricardo G. Poeira[1], Thomas P. Weiss[1], Renaud Leturcq[2], Alex Redinger[1] and Susanne Siebentritt[1]

[1]*Department of Physics and Materials Science, University of Luxembourg, 41 rue du Brill, Belvaux, L-4422 Luxembourg.*
[2]*Luxembourg Institute of Science and Technology—Materials Research and Technology Department, 41 rue du Brill, Belvaux, L-4422 Luxembourg.*
*Electronic mail: omar.ramirez@uni.lu*

**ABSTRACT**

Doping in the chalcopyrite Cu(In,Ga)Se$_2$ is determined by intrinsic point defects. In the ternary CuInSe$_2$, both N-type and P-type conductivity can be obtained depending on the growth conditions and stoichiometry: N-type is obtained when grown Cu-poor, Se-poor and alkali-free. CuGaSe$_2$, on the other hand, is found to be always a P-type semiconductor that seems to resist all kinds of N-type doping no matter whether it comes from native defects or extrinsic impurities. In this contribution, we study the N-to-P transition in Cu-poor Cu(In,Ga)Se$_2$ single crystals in dependence of the gallium content. Our results show that Cu(In,Ga)Se$_2$ can still be grown as an N-type semiconductor until the gallium content reaches the critical concentration of 15—19%, where the N-to-P transition occurs. Furthermore, trends in the Seebeck coefficient and activation energies extracted from temperature-dependent conductivity measurements, demonstrate that the carrier concentration drops by around two orders of magnitude near the transition concentration. Our proposed model explains the N-to-P transition based on the differences in formation energies of donor and acceptor defects caused by the addition of Gallium.

# I. INTRODUCTION

Despite sharing a similar electronic structure, one of the most puzzling differences between $CuInSe_2$ and $CuGaSe_2$ is the fact that the former can be intrinsically doped N or P-type while the latter is always a P-type semiconductor regardless of the growth conditions or deviations from molecularity and valence stoichiometry[1]. In the ternary $CuInSe_2$, there are three main parameters involved in determining its conductivity type: (I) the overall copper content[2], (II) the selenium pressure during or after growth[3] and (III) the presence of alkali metals[4-5]. In order to achieve N-type $CuInSe_2$, the sample must be alkali-free, have been grown under low Se pressure and be either Cu-poor or close-to-stoichiometric [2, 6]. Each of these conditions by its own have the capability to change the conductivity type from N to P, i.e.; having a Cu-rich composition, adding alkali metals either through a postdeposition treatment or from the soda lime glass[4], or annealing under a high selenium pressure[3], would result in a P-type $CuInSe_2$. Single crystals grown by metalorganic vapor phase epitaxy (MOVPE) comply with all three conditions to obtain N-type conductivity. As the selenium overpressure in MOVPE is considerably lower than in a co-evaporation[7], and the fact that selenium cannot be supplied during the cool down stage, it has been shown that alkali-free $CuInSe_2$ single crystals grown by MOPVE are always N-type regardless of the Se partial pressure used during growth as long as the composition is Cu-poor[4].

In the case of $CuGaSe_2$, Zunger et al. have performed extensive theoretical studies on the possibility to achieve N-type doping by extrinsic impurities such as H, Cd, Zn, Mg, Cl and other halogens [8-10], finding that none of them can effectively produce an N-type behavior. A recent study on the possibility to dope $CuGaSe_2$ N-type by hydrogen, concluded that the incorporation of H from an atomic source like a hydrogen plasma treatment could invert the conductivity from P to N-type[11], but no experimental evidence has been reported so far. The fact that $CuGaSe_2$ resists N-type doping has been pointed out as a part of a general trend in semiconductor families, where the wider-bandgap member exists often in only one doping type, like AlN when compared to GaN and InN, due to the so-called "doping limit rule"[9, 12].

The fact that $CuInSe_2$ can be grown as an N-type semiconductor under Cu-poor conditions and that $CuGaSe_2$ is always P-type, indicates that within Cu-poor $Cu(In,Ga)Se_2$ exists a transition point caused by the alloy with gallium. The aim of this contribution is to provide experimental evidence of such a transition and to understand the reasons for the change in the type of majority carriers. In order to carry out this investigation, several $Cu(In,Ga)Se_2$ single crystals of around 530 nm thickness were grown by metalorganic vapor phase epitaxy (MOVPE). Since the presence of alkalis and the Cu-content play a crucial role in determining the conductivity type, the copper content of all the $Cu_y(In,Ga)Se_2$ single crystals was restricted to $0.8 > y < 0.9$, while the presence of alkalis was suppressed by using undoped GaAs wafers as a substrate. Details of the heteroepitaxial growth, cross-section scanning electron microscope images, and a secondary-ion mass spectrometry analysis of selected samples, can be found in section S1 of the supplementary material.

## II.     METHODS

Photoluminescence spectra were obtained by exciting bare absorbers with a 660nm diode laser and the emitted photoluminescence collected into an InGaAs array spectrometer. All measurements were performed at room temperature and spectrally corrected employing a calibrated halogen lamp.

Temperature-dependent conductivity measurements were performed in a closed-cycle cryostat. Samples of 0.6 x 0.6 mm were prepared in the Van der Pauw configuration by evaporating triangular gold contacts with a thickness of 150 nm. For the Seebeck coefficient measurements, rectangular pieces of each sample were cleaved and mechanically pressed onto a home-made setup consisting of two copper pieces (one thermalized at room temperature and one heated). Details of the setup can be found elsewhere[13].

Energy-dispersive X-ray spectroscopy was carried out at 5kV and the L-line of all elements used for quantification. No traces of arsenic were detected at this acceleration voltage, ensuring that that gallium atomic percentage measured was not affected by the GaAs substrate.

XPS measurements were carried out using a hemispherical energy analyzer from Prevac (EA15) with a 2d detection system MCP/camera detector. The energy analyzer is assembled in the UHV analysis chamber from Scienta Omicron. A Kα x-ray source with a photon energy of 1486.6 eV was used in these measurements. The survey spectra was collected using the straight slit 2.5mm x 25mm, pass energy of 200eV and energy step of 0.192eV. The samples were mounted in the same sample holder using the same ground connection and then transferred without air exposure from a glovebox to the UHV XPS chamber under an inert gas transfer system.

### III. RESULTS AND DISCUSSION

Since the determination of the gallium content was of outmost importance for the purpose of this investigation, different techniques were used to measure the percentage of Ga present in each sample. Since an increase in bandgap due to the shift of the conduction band (CB) is expected for higher gallium contents[14-15], it is possible to approximate the Ga concentration from the optical bandgap dictated by the position of the maximum of the photoluminescence (PL) spectrum and the experimentally determined expression: $E_g = 1.01 + 0.626x - 0.167x(1-x)$ , where $x$ is the Ga content [15]. Fig. 1(a) displays the normalized PL spectra of all seven $CuIn_{1-x}Ga_xSe_2$ single crystals from which the gallium content was determined by the position of its maximum. Table I summarizes the samples' elemental composition determined by energy dispersive X-ray spectroscopy (EDX) and photoluminescence. The average of these quantities was rounded and used as the characteristic gallium content of each sample. Some samples were also analyzed by X-ray photoelectron spectroscopy (XPS) and Raman spectroscopy, which showed elemental compositions in agreement with the already determined by EDX and PL. Details of the XPS quantification and the Raman analysis can be found in section S2 of the supplementary information.

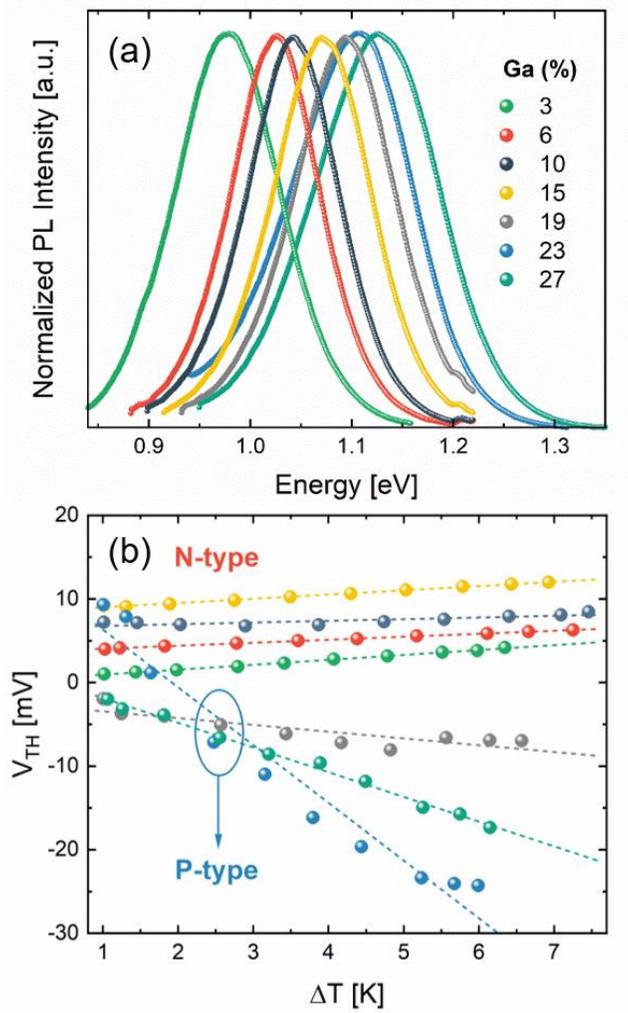

**FIG. 1.** Normalized photoluminescence spectra of all the $CuIn_{1-x}Ga_xSe_2$ samples showing the expected blue shift for higher gallium contents corresponding to the increase in bandgap (a). Seebeck coefficient determination (data shifted vertically for clarity) (b). The same color code applies to both figures.

**TABLE I**. Copper content, gallium percentage determined by different techniques and Seebeck coefficient of all the $Cu(In,Ga)Se_2$ single crystals investigated.

| Sample name | Cu/(In+Ga) | Ga % EDX 5kV | Ga % PL | Ga % Average | Ga % XPS | Seebeck coefficient [mV/K] |
|---|---|---|---|---|---|---|
| CIGSe298 | 0.86 | 5.2 | 0.0 | 3 | - | -0.58 ± 0.01 |
| CIGSe295 | 0.89 | 7.0 | 5.2 | 6 | 6.0 | -0.35 ± 0.01 |
| CIGSe296 | 0.89 | 10.4 | 8.8 | 10 | 10.0 | -0.20 ± 0.04 |
| CIGSe297 | 0.88 | 14.7 | 14.5 | 15 | 14.0 | -0.50 ± 0.01 |
| CIGSe299 | 0.88 | 19.5 | 18.8 | 19 | 20.0 | 0.80 ± 0.16 |
| CIGSe304 | 0.83 | 24.0 | 21.6 | 23 | - | 6.90 ± 0.54 |
| CIGSe305 | 0.82 | 29.2 | 25.2 | 27 | - | 2.95 ± 0.07 |

Once the gallium content of each sample was determined, the Seebeck coefficient ($S = -\Delta V_{TH}/\Delta$), which measures the thermoelectric voltage $V_{TH}$ generated as a response to an applied temperature difference $\Delta T$, was measured in order to investigate the type of majority carrier. Negative values of $S$ indicate that electrons are the majority carrier (N-type), and positive values, that conduction is carried by holes (P-type). The slope of the measured thermovoltages as a function of temperature gradients is equal to the Seebeck coefficient. Linear fits are presented in Fig. 1(b) for all investigated samples. These measurements confirmed the expected N-type character of the samples with the lowest gallium contents; however, a clear change in the thermoelectric behavior can be seen for gallium contents higher than 15%. First, the sign of the Seebeck coefficient changes from negative to positive (values listed in Table I), indicating that the majority carrier changes from electrons to holes (N to P transition). Besides of that, the magnitude of the Seebeck coefficient increases, which suggests that the Fermi level ($E_F$) has moved further away from the respective band edges, since the Seebeck coefficient is defined in terms of the semiconductor energy levels as[16]:

$$S_n = -\frac{k_B}{q}\left(\frac{5}{2} + r_e - \frac{E_F - E_C}{k_B T}\right)$$

$$S_p = \frac{k_B}{q}\left(\frac{5}{2} + r_h - \frac{E_V - E_F}{k_B T}\right)$$

For an N-type ($S_n$) and P-type ($S_p$) semiconductor. $k_B$ is the Boltzmann constant, $q$ the elementary charge, $r_{e,h}$ a term that depends on the carrier scattering mechanism, $E_F$ the Fermi level and $E_{V,C}$ the corresponding valence or conduction band energy. Values of the Seebeck coefficient for the strongly N-doped samples are in agreement with previous reports[17].

In order to corroborate the apparent decrease in carrier concentration after the transition from N to P, an analysis of the electrical conductivity ($\sigma$) at room temperature and its temperature-dependency for selected samples was carried out and the activation energy ($E_A$) determined from the Arrhenius plot in Fig. 2(a). All the N-type samples analyzed were found to be more conductive and have a lower activation energy (41-76 meV) than the P-type ones (296-431 meV). This difference in activation energy is probably due to the nature of the defects involved in the conductivity, first shallow donors and then deeper acceptors (as shallow acceptors are likely compensated). The low activation energy of the N-type samples, nonetheless, could partially be due to the influence of the thermally activated mobility, as activation energies in the range of 3-20 meV have been reported[4]. The increase in conductivity and activation energy measured, supports the Seebeck coefficient trend of a decrease in carrier concentration as the gallium content increases towards the N-to-P transition. In order to figure out whether the high activation energy was a characteristic of higher Ga contents only, a P-type sample but with a 7% gallium content was also analyzed. To achieve this, a potassium fluoride post-deposition treatment (KF-PDT) was performed on the N-type absorber in order to change the conductivity type to P. Evidence of the N-to-P type inversion in the exact same sample due to the KF-PDT can be found in Reference [18]. It is worth mentioning that the reported Seebeck coefficient of -0.346 mV/K for the N-type sample is almost the same as the one reported herein for a 6% Ga (-0.36 mV/K). Despite the low Ga content, the activation energy of the P-type KF-treated sample was still considerably larger than the $E_A$ of its N-type counterpart, suggesting that the high activation energies measured are actually a consequence of the majority carriers being holes (and their concentration), and not because of the increase in Ga content itself. A similar observation in pure $CuInSe_2$ has been reported for when the N-to-P transition is caused by the copper content [4].

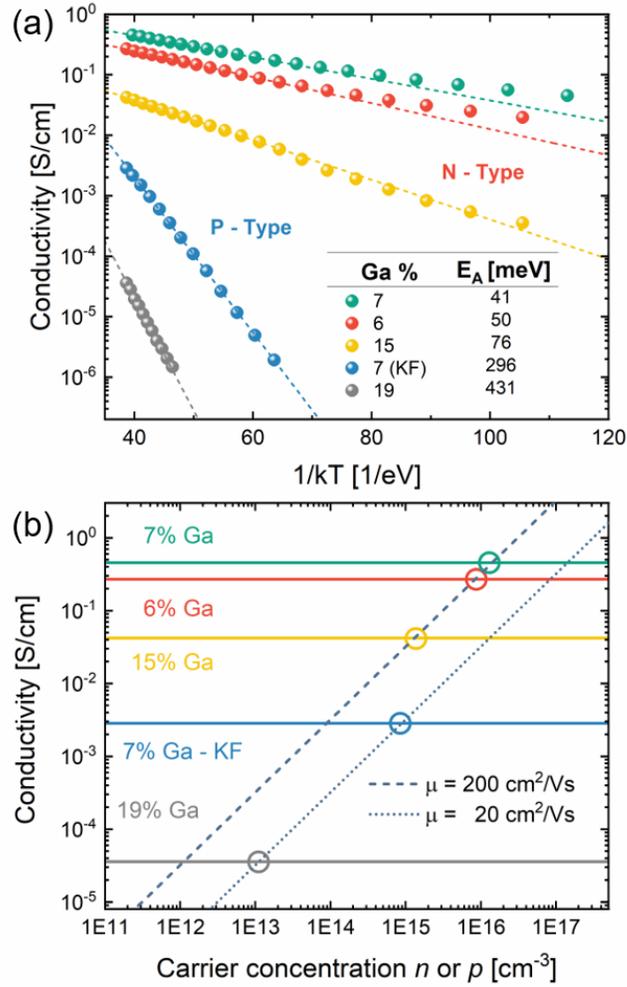

**FIG. 2.** Temperature-dependent conductivity measurements of N and P-type Cu(In,Ga)Se$_2$ single crystals (a). Best fits to the measured data are shown as dotted lines from which the activation energies were extracted. Determination of the carrier concentration from measured conductivity values for two different carrier mobilities (b).

From the measured conductivity values at room temperature and using p(n) >> n(p), we can estimate the carrier concentration as $p(n) = \sigma/q\mu_{h(e)}$, where $q$ is the elementary charge and $\mu_{h(e)}$ the hole (electron) mobility. Values of $\mu_e$ between 40 and 200 cm²/Vs measured by Hall have been reported for N-type CuInSe$_2$ single crystals grown by MOVPE with Cu contents ranging from 0.8 to 0.9[4], while values of $\mu_h$ in close-to-stoichiometry CuGaSe$_2$ samples, between 20 and 150 cm²/Vs[1, 19].

Carrier mobility in both ternaries is strongly dependent on the Cu content, as it has been demonstrated to increase towards Cu-rich compositions [4, 20]. Fig. 2(b) shows the estimated carrier concentration for the same set of N and P-type samples for two different mobility values. By taking $\mu_e$=200 cm$^2$/Vs, the carrier concentration of the N-type samples is estimated to be around $1.3\times10^{15}$ – $1.4\times10^{16}$ cm$^{-3}$; the P-type samples, on the other hand, are found to have carrier concentrations around two orders of magnitude lower for a hole mobility of $\mu_h$=20 cm$^2$/Vs. These electron and hole mobilities were chosen based on reported values for CuInSe$_2$ and CuGaSe$_2$ single crystals grown by MOVPE and with copper contents similar to the studied samples [4, 19]. Indeed, reported carrier concentrations for N-type single crystals agree with our findings on the magnitude of *n* being in the order of the $10^{16}$ cm$^{-3}$ [4, 21]. Carrier concentration in CuGaSe$_2$ on the other hand, has been reported to drastically decrease from around $10^{16}$ to $10^{14}$ cm$^{-3}$ below the stoichiometric point due to an increase in the degree of compensation [19, 22], in agreement with our estimated hole carrier concentrations of around $1.1\times10^{13}$ – $8.9\times10^{14}$ cm$^{-3}$.

As a way to confirm that the transition from N to P happens at gallium contents between 15 - 19%, X-ray photoelectron spectroscopy was used to analyze the change in binding energy of the constituent elements. Since a considerable change in conductivity happens at the N-to-P transition, a shift in the binding energy would be expected due to the electrostatic difference in the interaction between specimen and spectrometer and surface charge accumulation [23-24]. Similarly, XPS has been used to study doping changes in moderately doped N and P-type silicon[25]. Fig. 3 shows the Cu2p, In3d, Ga2p and Se3d binding energies of samples with gallium contents close to the N-to-P transition region. As can be seen, a considerable difference in the binding energy of approximately 1.6 eV was measured in the sample containing 19% gallium with respect to the other samples with lower gallium contents. Since the shift was not only in one specific element but in all the constituents, we attributed this to the change in conductivity due to the N-to-P transition rather than to a change in chemical environment of a specific element.

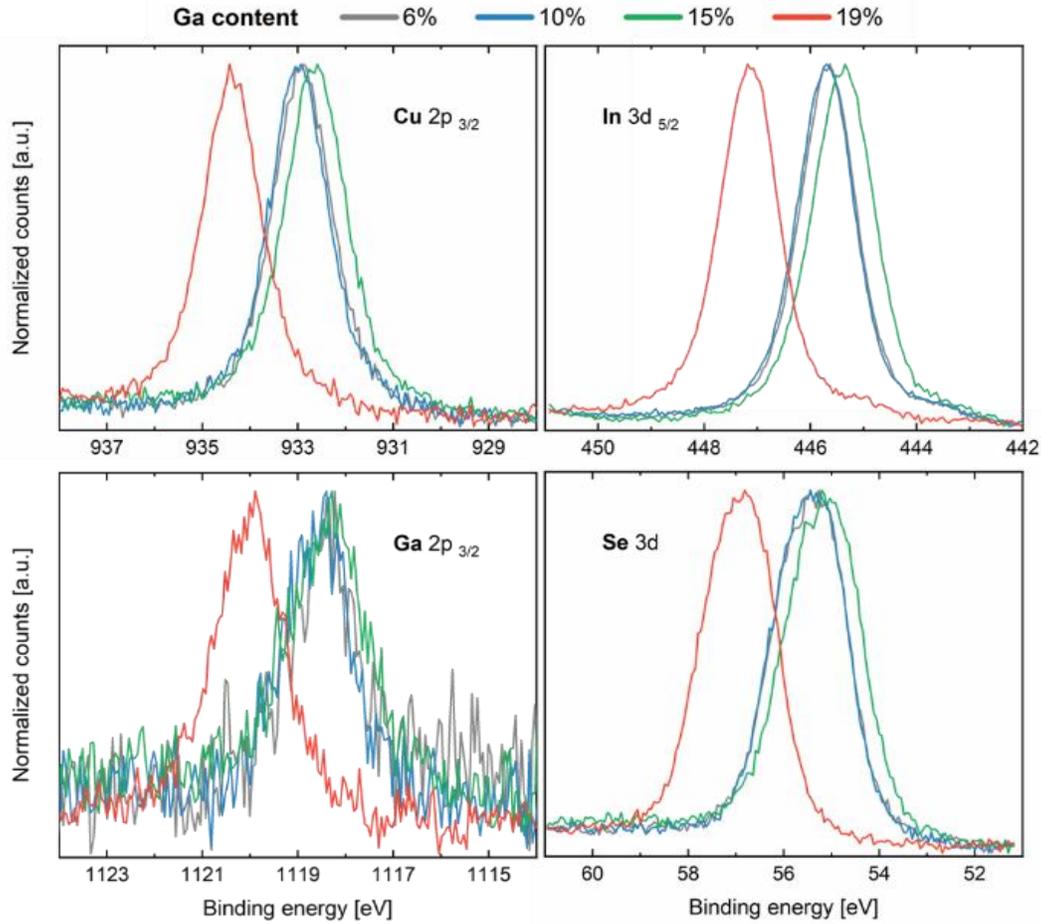

**FIG. 3.** Normalized In3d peaks of samples with different Ga content around the critical gallium concentration of 15 to 19%. The shift in binding energy is associated to the change in conductivity type from N-to-P.

So far, we have demonstrated that the conductivity type due to the addition of gallium to $CuInSe_2$ changes from N to P-type when the gallium content is between 15 and 19%, but we have not addressed the explanation behind this observation. The electronic structure of both $CuInSe_2$ and $CuGaSe_2$ ternary compounds is similar: at least two shallow acceptors (A1 - A2) located at 40 - 60 meV away from the valance band (VB) in $CuInSe_2$ and 60 – 100 meV in $CuGaSe_2$; and a shallow donor (D1) at 10meV below the conduction band for both cases [26-27]. Theoretical studies[28-30] have assigned the origin of these electronic states to intrinsic point defects in the chalcopyrite crystal structure: A1→Cu vacancies ($V_{Cu}$), A2→Cu on an In/Ga site ($Cu_{III}$) and D1→ Cu interstitial ($Cu_i$) or In on a Cu site (In$_{Cu}$). For a more in-depth overview of defects in Cu(In,Ga)Se$_2$, the reader is referred to the review by Spindler et al[27]. Experimentally, neutron powder diffraction has been used to determine defect concentrations of alkali-free CuInSe$_2$, finding that the N-type character of Cu-poor samples is given by the shallow donor In$_{Cu}$ substitutional defect[31], which goes along with its theoretically calculated low formation energy[28, 32].

The probability of the formation of defects depends on the chemical potential (Δμ) of the constitutional elements, which in turn depends on the crystal growth conditions. In the work of Pohl et al., the formation energies of intrinsic point defects in both CuInSe$_2$ and CuGaSe$_2$ were calculated for different chemical potentials[28]. For selenium-poor, Cu and In-rich conditions (Δμ$_{In}$ = -0.2 eV and Δμ$_{Cu}$ = 0 eV, point D in Fig.1-up [28]) the intrinsic Fermi level in CuInSe$_2$ was found to be closer to the conduction band. On the contrary, under the same conditions (most Se poor point at Δμ$_{Ga}$ = -0.3 eV, point D in Fig.1-down [28]), the intrinsic Fermi level in CuGaSe$_2$ was found below midgap. Thus, the theory also predicts an N-to-P transition due to the addition of Ga. Therefore, we analyzed the trends in formation energy of the previously described donors and acceptors in order to understand what happens at the critical gallium concentration of 15-19%. The validity of this analysis resides on the fact that changes in the chemical potential would result in a shift of all formation energies, but would not affect the observed trends.

Since the formation energy of defects depends on the position of the Fermi level, we firstly estimate E$_F$ using the carrier concentrations previously obtained and literature values for the effective density of states[33] (details of the calculations can be found in section S3 of the SI). From this, we obtained that in the case of the N-type sample with 15% gallium, E$_F$ is approximately 160 meV away from the CB and in the case of the P-type sample with 19%, at 360 meV from the VB. Fig. 4(b) shows the position of the calculated Fermi level for these and other selected samples. With this information, we proceeded to analyze the change in formation energies of A1, A2 and D1 in two situations: (A) from an N-type CuInSe$_2$ (E$_F$ = 200 meV from CB) to an N-type CuGaSe$_2$ (E$_F$ = 400 meV from CB) and (B) from an N-type CuInSe$_2$ (E$_F$ = 200 meV from CB) to a P-type CuGaSe$_2$ (E$_F$ = 400 meV from VB). The values of the formation energies were taken from reference [28] and the two cases are illustrated in Fig. 4(a) (more details in section S4 of the supplementary material). By analyzing case A, it is possible to make the following deductions: With the increase of gallium, (I) the formation energy of both possible donors increases, (II) the formation energy of both acceptors decreases and (III) the formation energy of A1 approaches zero. As a result, N-type CuGaSe$_2$ becomes very unlikely. For case B (which represents what has been experimentally observed), we observe the opposite trend. Interestingly, the defect with the lowest formation energy changes depending on the gallium content. For Ga/(Ga+In) < 0.5, the acceptor V$_{Cu}$ dominates, but in higher gallium contents, the donor III$_{Cu}$ has the lowest formation energy, which would go along with reports of Cu-poor CuGaSe$_2$ being strongly self-compensated[34].

From the previous analysis, we can explain our experimental results as follows: When gallium is introduced to N-type CuInSe$_2$, the Fermi level starts to move away from the conduction band, because due to the lower formation energy of acceptor-type defects, more acceptors are formed. If the conductivity remained N-type (case A), further addition of gallium would result in the spontaneous formation of V$_{Cu}$ (formation energy approaching zero), which would move the Fermi level below midgap making the material P-type, which is always the case in CuGaSe$_2$. When the gallium content reaches the critical concentration of 15—19%, as the formation of acceptor defects becomes more energetically favorable, the acceptor density (N$_A$) overpasses the donors (N$_D$) resulting in the material changing to P-type. In this situation (case B), further addition of gallium would result in an increased degree of compensation ($K = N_D/N_A$ for a P-type semiconductor) as the formation energy of donor-like defects decreases. As an unavoidable consequence of the increased degree of compensation, stronger electrostatic potential fluctuations would be expected as the gallium content increases. Experimental evidence of this can be found in literature, where the magnitude of these fluctuations (denoted as γ) in CuGaSe$_2$ was found to be greater than in CuInSe$_2$ for copper contents around 0.9 ($\gamma_{CGSe}$ = 29-36 meV and $\gamma_{CISe}$ =15-28 meV)[35]. Our own studies in stoichiometric alkali-free Cu(In,Ga)Se$_2$ single crystals[36], also support this implication as we have observed experimental evidence of higher Urbach energies caused by electrostatic potential fluctuations in Cu(In,Ga)Se$_2$ than in CuInSe$_2$.

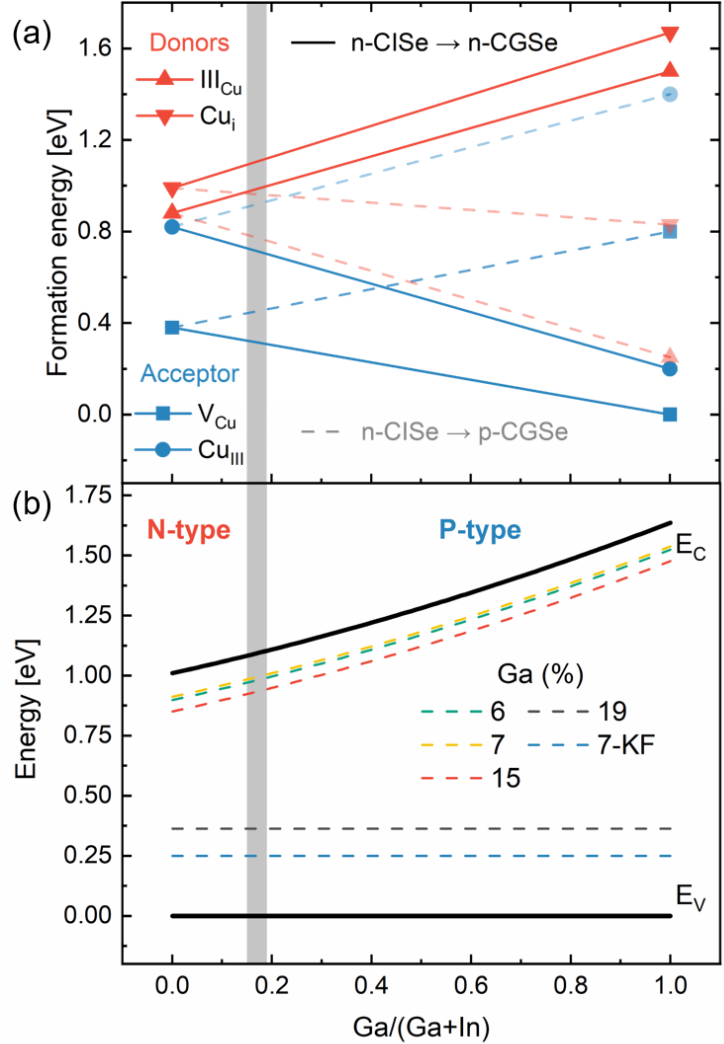

**FIG. 4.** Formation energy of the two possible donor defects III$_{Cu}$ and Cu$_i$ (red) and the two shallow acceptors A1 and A2 (blue). The points for CuInSe$_2$ correspond to a Fermi level 200 meV away from the conduction band (N-type) while two sets of points are displayed for CuGaSe$_2$, assuming an N-type (solid line) and a P-type (dotted line), in both cases the Fermi level is assumed to be 400 meV away from the bands (a). All formation energies were taken from the work of Pohl et al.[28] Band diagram showing the calculated Fermi level for N and P-type samples with different gallium contents (b). The energy of the valance band is assumed constant and the bandgap calculated according to the equation given in the text.

## IV. CONCLUSION

In summary, we studied the N-to-P transition in Cu-poor CuInSe$_2$ caused by the alloy with gallium. Our results demonstrated that Cu(In,Ga)Se$_2$ can be intrinsically grown as an N-type semiconductor as long as the gallium content is below the critical concentration of 15—19%. The transition from N to P-type was confirmed by the change in the sample's thermoelectric behavior and the shift in binding energy measured by XPS. Furthermore, by measuring electrical conductivity and taking an estimated mobility, we found that the carrier concentration has a decreasing trend towards the N-to-P transition region, dropping around two orders of magnitude (from n ≈ 1.3x10$^{15}$ – 1.4x10$^{16}$ cm$^{-3}$ to p ≈ 1.1x10$^{13}$ – 8.9x10$^{14}$ cm$^{-3}$) when the material becomes P-type. By analyzing the trends in formation energy of donor and acceptor-like defects, we concluded that the N-to-P transition due to the addition of gallium is caused by: (I) the more energetically favored formation of acceptor-like defects as the formation energy of acceptor states decreases and of donor states increases and (II) the fact that a Fermi level above midgap results in the instant formation of V$_{Cu}$, preventing N-type doping. With this contribution, we aimed at providing experimental evidence and addressing the long-standing discussion in the chalcopyrite community on the possibility to grow N-type Cu(In,Ga)Se$_2$.

**SUPPLEMENTARY MATERIAL**

See the supplementary material for the details on the heteroepitaxial growth, XPS, Raman spectroscopy, Fermi level calculations and formation energies determination.

**ACKNOWLEDGEMENTS**

This work was supported by the Luxembourgish Fond National de la Recherche in the framework of the projects C17/MS/11696002 GRISC, PRIDE17/12246511/PACE and 11244141 SUNSPOT. The authors acknowledge and thank Dr. Nathalie Valle and Brahime El Adib from the Luxembourg Institute of Science and Technology for all his support regarding the SIMS measurements. The authors also thank Dr. Florian Werner for his contribution with the conductivity measurements of selected samples and the fruitful discussions, as well as Thomas Schuler, Bernd Uder and Ulrich Siegel for their technical support. For the purpose of open access, the author has applied a Creative Commons Attribution 4.0 International (CC BY 4.0) license to any Author Accepted Manuscript version arising from this submission.

**AUTHOR DECLARATIONS**

**Conflict of Interest**

The authors have no conflicts to disclose.

**DATA AVAILABILITY**

The data that support the findings of this study are openly available in Zenodo at https://doi.org/10.5281/zenodo.6362649 as well as from the corresponding author upon reasonable request.

# REFERENCES

[1] L. Mandel, R. D. Tomlinson, M. J. Hampshire, H. Neumann, *Solid State Communications* **1979**, 32, 201.
[2] H. Neumann, R. D. Tomlinson, *Solar Cells* **1990**, 28, 301.
[3] J. Parkes, R. D. Tomlinson, M. J. Hampshire, *J. Cryst. Growth* **1973**, 20, 315.
[4] F. Werner, D. Colombara, M. Melchiorre, N. Valle, B. El Adib, C. Spindler, S. Siebentritt, *J. Appl. Phys.* **2016**, 119, 173103.
[5] D. J. Schroeder, A. A. Rockett, *J. Appl. Phys.* **1997**, 82, 4982.
[6] R. Noufi, R. Axton, C. Herrington, S. K. Deb, *Appl. Phys. Lett.* **1984**, 45, 668.
[7] F. Babbe, H. Elanzeery, M. H. Wolter, K. Santhosh, S. Siebentritt, *J. Phys. Condens. Matter* **2019**, 31, 425702.
[8] C. Persson, Y.-J. Zhao, S. Lany, A. Zunger, *Phys. Rev. B* **2005**, 72, 035211.
[9] Y.-J. Zhao, C. Persson, S. Lany, A. Zunger, *Appl. Phys. Lett.* **2004**, 85, 5860.
[10] Ç. Kılıç, A. Zunger, *Phys. Rev. B* **2003**, 68, 075201.
[11] M. Han, P. Deák, Z. Zeng, T. Frauenheim, *Phys. Rev. Appl.* **2021**, 15, 044021.
[12] S. B. Zhang, S. H. Wei, A. Zunger, *Phys. Rev. Lett.* **2000**, 84, 1232.
[13] P. Lunca-Popa, J. Afonso, P. Grysan, J. Crêpellière, R. Leturcq, D. Lenoble, *Sci. Rep.* **2018**, 8, 7216.
[14] S.-H. Wei, S. B. Zhang, A. Zunger, *Appl. Phys. Lett.* **1998**, 72, 3199.
[15] M. I. Alonso, M. Garriga, C. A. Durante Rincón, E. Hernández, M. León, *Appl. Phys. A* **2002**, 74, 659.
[16] C. Herring, *Physical Review* **1954**, 96, 1163.
[17] S. Endo, T. Irie, H. Nakanishi, *Solar Cells* **1986**, 16, 1.
[18] O. Ramírez, M. Bertrand, A. Debot, D. Siopa, N. Valle, J. Schmauch, M. Melchiorre, S. Siebentritt, *Sol. RRL.* **2021**, 5, 2000727.
[19] A. Gerhard, W. Harneit, S. Brehme, A. Bauknecht, U. Fiedeler, M. C. Lux-Steiner, S. Siebentritt, *Thin Solid Films* **2001**, 387, 67.
[20] S. Siebentritt, S. Schuler, *Journal of Physics and Chemistry of Solids* **2003**, 64, 1621.
[21] S. M. Wasim, *Solar Cells* **1986**, 16, 289.
[22] S. Siebentritt, L. Gütay, D. Regesch, Y. Aida, V. Deprédurand, *Sol. Energy Mater. Sol. Cells* **2013**, 119, 18.
[23] D. R. Baer, K. Artyushkova, H. Cohen, C. D. Easton, M. Engelhard, T. R. Gengenbach, G. Greczynski, P. Mack, D. J. Morgan, A. Roberts, *J. Vac. Sci. Technol. A* **2020**, 38, 031204.
[24] D. R. Baer, C. F. Windisch Jr, M. H. Engelhard, K. R. Zavadil, *Journal of Surface Analysis* **2002**, 9, 396.
[25] H. Sezen, S. Suzer, *The Journal of Chemical Physics* **2011**, 135, 141102.
[26] S. Siebentritt, M. Igalson, C. Persson, S. Lany, *Prog Photovolt.* **2010**, 18, 390.
[27] C. Spindler, F. Babbe, M. H. Wolter, F. Ehré, K. Santhosh, P. Hilgert, F. Werner, S. Siebentritt, *Phys. Rev. Mater.* **2019**, 3, 090302.
[28] J. Pohl, K. Albe, *Phys. Rev. B* **2013**, 87, 245203.
[29] M. Malitckaya, H.-P. Komsa, V. Havu, M. J. Puska, *Advanced Electronic Materials* **2017**, 3, 1600353.
[30] L. E. Oikkonen, M. G. Ganchenkova, A. P. Seitsonen, R. M. Nieminen, *J. Phys. Condens. Matter* **2014**, 26, 345501.
[31] C. Stephan, S. Schorr, M. Tovar, H.-W. Schock, *Appl. Phys. Lett.* **2011**, 98, 091906.
[32] J. Bekaert, R. Saniz, B. Partoens, D. Lamoen, *Phys. Chem. Chem. Phys.* **2014**, 16, 22299.
[33] J. L. Gray, R. Schwartz, Y. J. Lee, *Purdue University - ECE Technical Reports* **1994**, 173.
[34] S. Schuler, S. Siebentritt, S. Nishiwaki, N. Rega, J. Beckmann, S. Brehme, M. C. Lux-Steiner, *Phys. Rev. B* **2004**, 69, 045210.

[35]   S. Siebentritt, N. Papathanasiou, M. C. Lux-Steiner, *Physica B: Condensed Matter* **2006**, 376-377, 831.
[36]   O. Ramírez, J. Nishinaga, M. Melchiorre, S. Siebentritt, *In preparation* **2022**.

# Supplementary Information

# How much gallium do we need for a p-type Cu(In,Ga)Se$_2$?

Omar Ramírez[1*], Evandro Martin Lanzoni[1], Ricardo G. Poeira[1], Thomas P. Weiss[1], Renaud Leturcq[2], Alex Redinger[1] and Susanne Siebentritt[1]

[1]Department of Physics and Materials Science, University of Luxembourg, 41 rue du Brill, Belvaux, L-4422 Luxembourg.
[2]Luxembourg Institute of Science and Technology—Materials Research and Technology Department, 41 rue du Brill, Belvaux, L-4422 Luxembourg.
*Electronic mail: omar.ramirez@uni.lu

## S1. Heteroepitaxial growth

The CuIn$_{1-x}$Ga$_x$Se$_2$ single crystals used in these experiments were grown by metalorganic vapor phase epitaxy (MOVPE) on semi-insulating 500μm thick (100)-oriented GaAs wafers. The metalorganic precursors utilized were cyclopentadienyl-coppertriethyl phosphine (CpCuTEP), trimethylindium (TMIn), triethylgallium (TEGa) and diisopropylselenide (DiPSe). The reactor growth conditions (T = 520°C and P = 90mbar) were kept the same for all processes.

In order to achieve the desired gallium and copper contents, a 2-step growth process was implemented. Firstly, a CuIn$_{0.6}$Ga$_{0.4}$Se$_2$ layer is grown, followed by a second layer of CuInSe$_2$ with a similar Cu content ($P_{Cu}/(P_{In} + P_{Ga})$ = 2.22–2.24, where $P$ is the partial pressure of the corresponding metalorganic precursor ). By changing the duration of the first step, the gallium content can be finely tuned. The duration of the second step is adjusted in such a way that the final growth time is always 8 hours. The partial pressure of DiPSe was adjusted in order to have a similar selenium overpressure of $P_{Se}/P_{Metal} \approx 25$ in both steps. Since the growth rate of the CuIn$_{0.6}$Ga$_{0.4}$Se$_2$ and CuInSe$_2$ layers is similar, no evident variation in the thickness of the samples (~530nm) was observed, as can be seen in Figure S1.

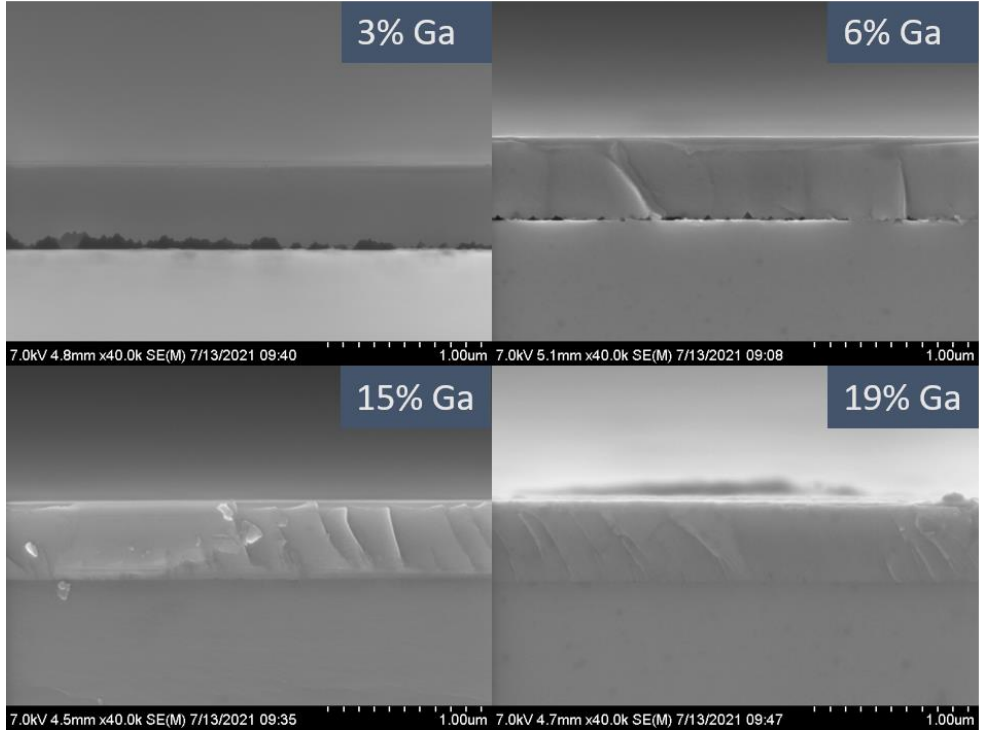

**FIG. S1.** SEM cross-section images of four different samples with gallium contents ranging from 3 to 19%. The thickness of all samples is around 530 nm. Notice the improvement at the interface with GaAs, due to a reduced lattice mismatch, as the gallium content increases.

A secondary-ion mass spectrometry (SIMS) analysis of the samples with 6 and 19% was carried out in order to investigate the gallium gradients caused by the diffusion of gallium from the first step of the growth process towards the pure $CuInSe_2$ of the second step, as well as the presence of sodium and potassium. As can be seen from Figure S2 (a), a noticeable and comparable gallium gradient towards the back is present in both samples, which discards a possible influence on the obtained results due to the gallium gradient as all samples seem to be equally affected. Furthermore, the GGI signal of the sample with 19% gallium is 3.17 times higher than the 6% Ga content within the first 1000 seconds of sputtering (equivalent to roughly 150 nm), which goes along the previously determined gallium content difference from other techniques.

Regarding alkali metals, Figure 2(b) and (c) show the SIMS depth profiles of $^{23}Na$ and $^{39}K$, respectively. In the case of sodium, an apparent segregation seems to occur at the GaAs interface. Nevertheless, since the width of the sodium peak is proportional to the CIGSe/GaAs interface, i.e., a sharp peak for the smooth interface with the sample with 19% Ga and a broad peak for the rough interface with 6% Ga, we attribute the apparent increase in $^{23}$Na intensity to the change in matrix composition and the so-called SIMS matrix effect. A similar sodium peak at the GaAs interface has been showed by Colombara et.al. in sodium-free CuInSe2 single crystals[1]. In the case of potassium, no traces were found along the full depth of the absorber layer.

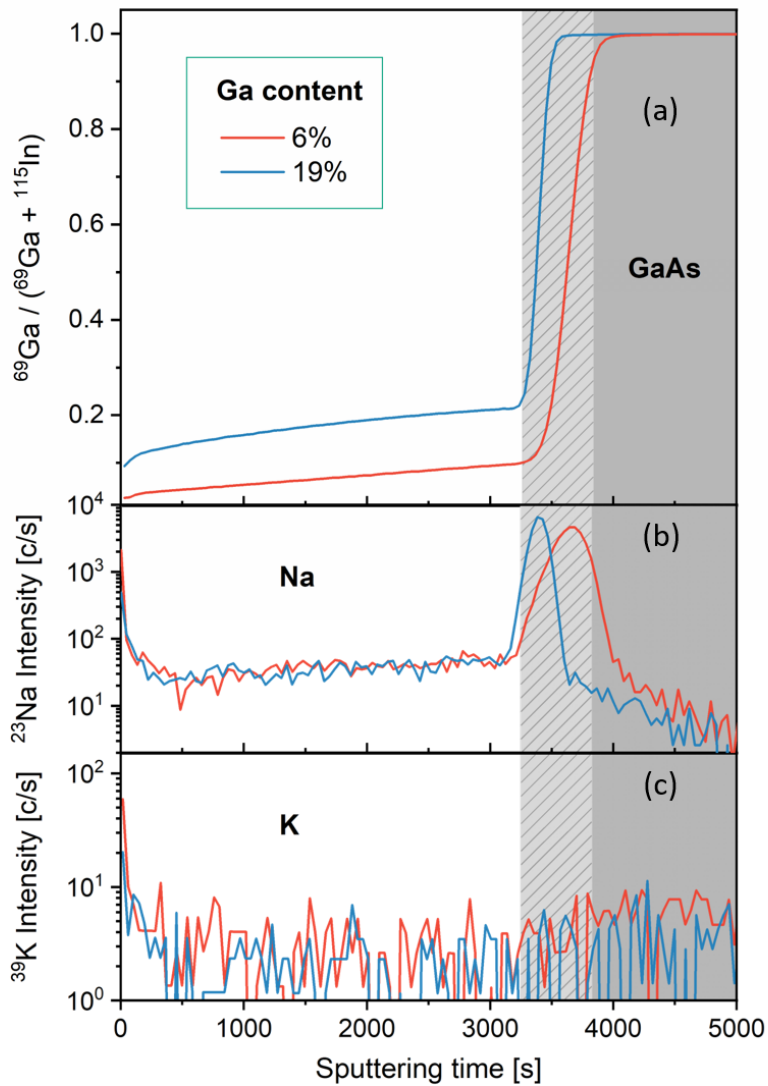

**FIG. S2.** Composition profiles obtained by secondary ion mass spectrometry of two samples used in this study containing 6 and 19% gallium. Ga/(Ga+In) signal ratio showing the resulting gallium gradient towards the GaAs interface (a), as well as the sodium (b) and potassium (c) profiles. The gray marked areas denote the interface (light) and the GaAs wafer (dark).

## S2. X-ray photoelectron and Raman spectroscopy

### S2.1 X-ray photoelectron spectroscopy

Quantification of the constituent elements was done from the general survey as well as from the Ga3d-In4d region (Fig. S3) as described in reference [2]. The reported values in the main manuscript of the Ga/(Ga+In) ratio are an average of the two methods, which are summarized in table S1.

**Table S1**. Ga and In atomic percentages as well as the Ga/(Ga+In) ratio calculated from XPS.

| Sample | From survey | | | From Ga3d-In4d fits | | | | | Average GGI |
|---|---|---|---|---|---|---|---|---|---|
| | In [at. %] | Ga [at. %] | GGI | Ga 3d 5/2 [at. %] | Ga 3d 3/2 [at. %] | In 4d 5/2 [at. %] | In 4d 3/2 [at. %] | GGI | |
| CIGSe295 | 40.97 | 2.05 | 0.05 | 3.95 | 3.83 | 47.15 | 45.07 | 0.07 | **0.06** |
| CIGSe296 | 41.00 | 3.38 | 0.08 | 6.16 | 5.98 | 44.92 | 42.94 | 0.12 | **0.10** |
| CIGSe297 | 37.85 | 5.12 | 0.12 | 8.01 | 7.77 | 43.06 | 41.16 | 0.16 | **0.14** |
| CIGSe299 | 37.21 | 7.89 | 0.17 | 11.77 | 11.42 | 39.27 | 37.54 | 0.23 | **0.20** |

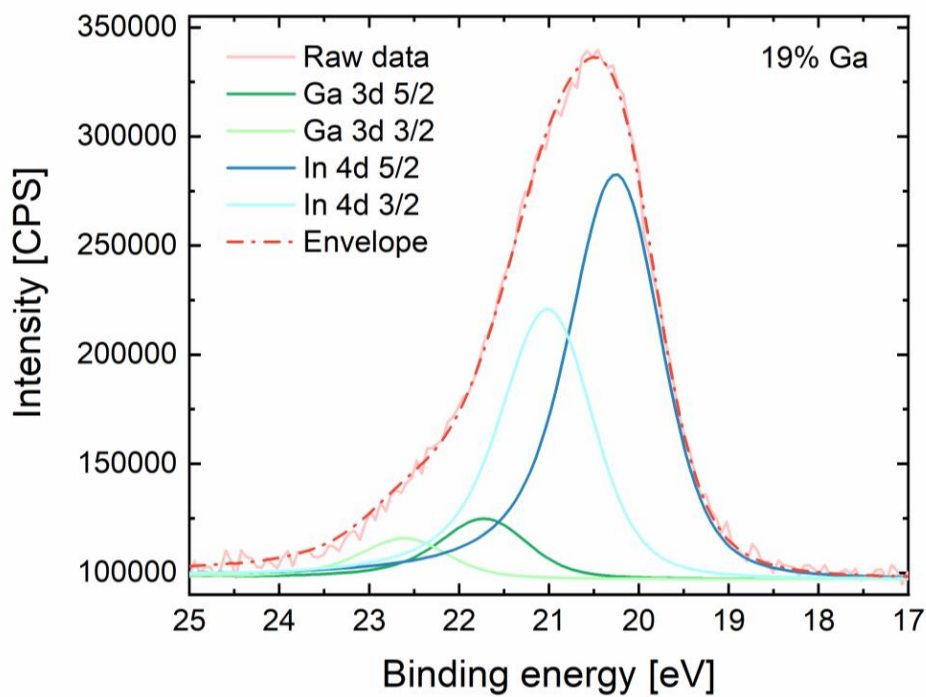

**FIG. S3.** X-ray photoelectron spectroscopy survey of four samples with different gallium content (a). Example of the fittings to the Ga3d-In4d region used to quantify the Ga/(Ga+In) ratio.

## S2.2 Raman spectroscopy

Raman measurements were taken with a Renishaw inVia micro spectrometer at room temperature with a 532nm laser excitation focused on the sample with a 50× objective lens and a numerical aperture of 0.5 in combination with a 2400 lines/mm grating. In order to approximate the gallium content from the measured Raman spectra, the data was fitted with a Lorentzian function from which the center $x_c$ was determined. Cu(In,Ga)Se$_2$ has its predominant A$_1$ vibrational mode ranging from 175 cm$^{-1}$ in CuInSe$_2$[3-4] to 184 cm$^{-1}$ in CuGaSe$_2$[5-7]. The gallium content was estimated from a linear interpolation and the results summarized in table S2. Most samples show good agreement with the average Ga contents determined from EDX and PL. The results of these measurements further confirm that the gallium content of the N-to-P transition is below 20%.

**Table S2**. Raman peak position determined from the Lorentzian fits and estimated gallium content.

|  | CIGSe295 | CIGSe296 | CIGSe297 | CIGSe299 |
|---|---|---|---|---|
| $x_c$ [cm$^{-1}$] | 175.48 ± 0.031 | 176.03 ± 0.025 | 176.04 ± 0.024 | 176.62 ± 0.027 |
| Ga [%] | 5.3 | 11.4 | 11.6 | 18.0 |

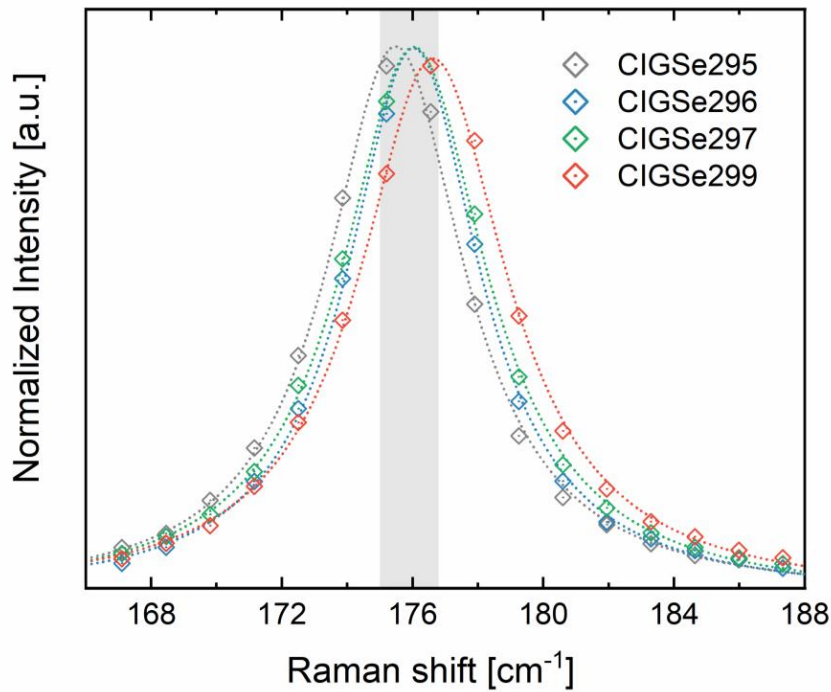

**FIG. S4.** Raman spectra of four samples with different gallium content. Dotted lines represent Lorentzian fits to the measured data from which the position of the A1 mode was determined. The shadowed area represent the expected A1 peak position for samples containing up to 20% gallium.

## S3. Fermi level determination

The calculated carrier concentrations obtained from measured conductivity values at room temperature are listed below:

| Ga content (%) | σ (S cm^-1) | Carrier | Assumed Mobility (cm^2 / Vs) | Inferred Carrier concentration (cm^-3) |
|---|---|---|---|---|
| 6 | 2.67E-01 | n | 200 | 8.3E+15 |
| 7 | 4.55E-01 | n | 200 | 1.4E+16 |
| 15 | 4.19E-02 | n | 200 | 1.3E+15 |
| 7 (KF) | 2.85E-03 | p | 20 | 8.9E+14 |
| 19 | 3.59E-05 | p | 20 | 1.1E+13 |

Using these values of n and p and reported[8] effective density of states of the valance band ($N_V$ = 1.5x10$^{19}$ cm$^{-3}$) and conduction band ($N_C$ = 6.6x10$^{17}$ cm$^{-3}$), the Fermi level was determined from the expressions:

$$n = N_C \exp\left(-\frac{E_C - E_F}{kT}\right)$$

$$p = N_V \exp\left(-\frac{E_F - E_V}{kT}\right)$$

Values of $E_F$ with respect to the closest band are:

| Ga content (%) | $N_C/N_V$ (cm^-3) | $E_F$ (meV) |
|---|---|---|
| 6 | 6.6E17 | 112.4 |
| 7 | 6.6E17 | 98.7 |
| 15 | 6.6E17 | 159.9 |
| 7 (KF) | 1.5E19 | 250.6 |
| 19 | 1.5E19 | 362.6 |

## S4. Formation energies determination

Firstly, the values of the formation energies for CuInSe$_2$ and CuGaSe$_2$, assuming that the Fermi level is located at the valance band and the conduction band, were taken from the work of Pohl et.al. and are displayed in Figure S5 [9]. For simplicity, the formation energies were taken at the Fermi level of interest regardless of the defect equilibrium charge state. For N-type CuInSe$_2$, values of the formation energies were taken from a linear interpolation at a Fermi energy of 0.2 eV. For CuGaSe$_2$, values were taken at 0.4 eV away from the valance and conduction band.

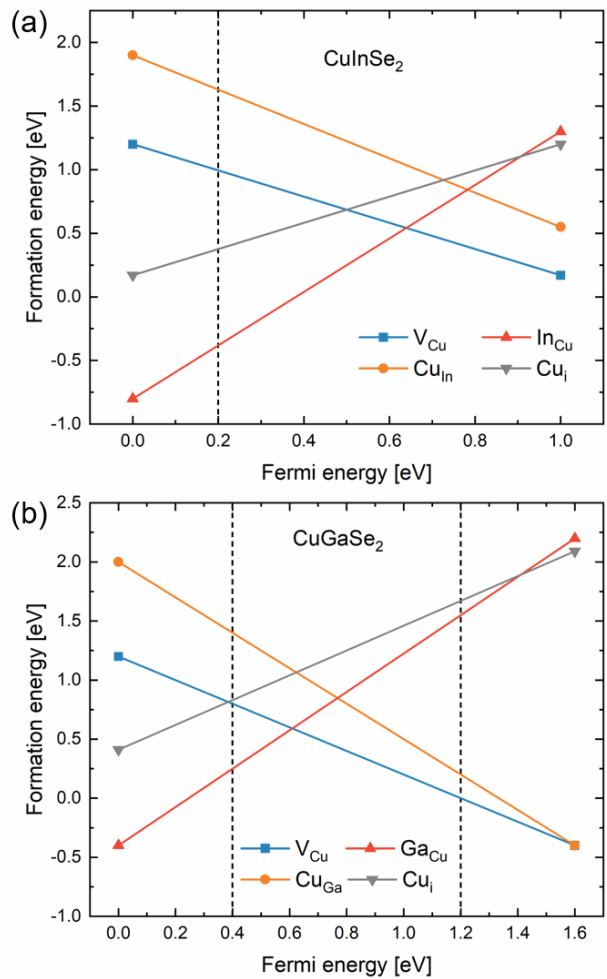

**FIG. S5.** Intrinsic defects' formation energy in CuInSe$_2$ (a) and CuGaSe$_2$ (b). Dashed lines intercept the formation energies at the Fermi levels considered in the text.

# References

[1]  D. Colombara, F. Werner, T. Schwarz, I. Cañero Infante, Y. Fleming, N. Valle, C. Spindler, E. Vacchieri, G. Rey, M. Guennou, M. Bouttemy, A. G. Manjón, I. Peral Alonso, M. Melchiorre, B. El Adib, B. Gault, D. Raabe, P. J. Dale, S. Siebentritt, *Nat. Commun.* **2018**, 9, 826.
[2]  S. Béchu, A. Loubat, M. Bouttemy, J. Vigneron, J.-L. Gentner, A. Etcheberry, *Thin Solid Films* **2019**, 669, 425.
[3]  H. Tanino, T. Maeda, H. Fujikake, H. Nakanishi, S. Endo, T. Irie, *Phys. Rev. B* **1992**, 45, 13323.
[4]  B. J. Stanbery, S. Kincal, S. Kim, C. H. Chang, S. P. Ahrenkiel, G. Lippold, H. Neumann, T. J. Anderson, O. D. Crisalle, *J. Appl. Phys.* **2002**, 91, 3598.
[5]  C. Xue, D. Papadimitriou, Y. S. Raptis, N. Esser, W. Richter, S. Siebentritt, M. C. Lux-Steiner, *J. Appl. Phys.* **2003**, 94, 4341.
[6]  C. Rincón, F. J. Ramírez, *J. Appl. Phys.* **1992**, 72, 4321.
[7]  D. Papadimitriou, N. Esser, C. Xue, *physica status solidi (b)* **2005**, 242, 2633.
[8]  J. L. Gray, R. Schwartz, Y. J. Lee, *Purdue University - ECE Technical Reports* **1994**, 173.
[9]  J. Pohl, K. Albe, *Phys. Rev. B* **2013**, 87, 245203.